\documentclass[fleqn,10pt]{wlscirep}
\usepackage[utf8]{inputenc}
\usepackage[T1]{fontenc}

\usepackage{graphicx,epsfig}% Include figure files
\usepackage{amssymb,amsmath}
\usepackage{soul}

\DeclareMathAlphabet{\mathcal}{OMS}{cmsy}{m}{n}

\def\cQ{{\mathcal Q}}

\def\cC{{\mathcal C}}
\def\cT{{\mathcal T}}

\usepackage{xcolor}

\newcommand{\req}[1]{Eq.~(\ref{#1})}

\newcommand{\fig}[1]{Fig.~\ref{#1}}

\newcommand{\red}{\textcolor{black}}

\newcommand{\cut}[1]{{}}

\newcommand{\vs}{{\vec{s}}}
\newcommand{\vP}{{\vec{P}}}
\newcommand{\pg}{{P_{g}}}
\newcommand{\bs}{{\beta_{s}}}

\title{Complete Realization of Energy Landscapes and Non-equilibrium Trapping Dynamics in Small Spin Glass and Optimization Problem}

\author[1,2]{Ho Fai Po}
\author[1,*]{Chi Ho Yeung}
\affil[1]{Department of Science and Environmental Studies, The Education University of Hong Kong, 10 Lo Ping Road, Hong Kong, China}
\affil[2]{Department of Mathematics, Aston University, B4 7ET, Birmingham, United Kingdom}

\affil[*]{chyeung@eduhk.hk}

\begin{abstract}
Energy landscapes are high-dimensional surfaces underlie all physical systems, which determine crucially the energetic and behavioral dependence of the systems on variable configurations, but are difficult to be analyzed due to their high-dimensional nature. Here we introduce an approach to reveal for the complete energy landscapes of spin glasses and Boolean satisfiability problems \red{with a small system size}, and unravels their non-equilibrium dynamics at an arbitrary temperature for an arbitrarily long time. Remarkably, our results show that it can be less likely for the system to attain ground states when temperature decreases, due to trapping in individual local minima, which ceases at a different time, leading to multiple abrupt jumps in the ground-state probability. For large systems, we introduce a variant approach to extract partially the energy landscapes and observe both \red{semi-}analytically and in simulations similar phenomena. This work introduces new methodology to unravel the energy landscapes and non-equilibrium dynamics of glassy systems, and provides us with a clear, complete and new physical picture on their long-time behaviors inaccessible by existing approaches.
\end{abstract}
\begin{document}

\flushbottom
\maketitle
\thispagestyle{empty}

%\noindent Please note: Abbreviations should be introduced at the first mention in the main text – no abbreviations lists. Suggested structure of main text (not enforced) is provided below.

\section*{Introduction}
Energy landscapes of physical systems are high-dimensional surfaces representing the dependence of system energy on variable configurations. Similarly, cost or fitness landscapes can be defined for optimization problems. Their characteristics determine crucially the emergent behavior of these systems. For instance, spin glasses and ferromagnetic spin systems are believed to have energy landscapes with and without a large number of local minima respectively~\cite{mezard1987spin, nishimori2001statistical}; a similar analogy is made with the algorithmic-hard and -easy phases of combinatorial optimization problems~\cite{krzakala2007gibbs}. A way to unravel and analyze the complete energy landscape is thus crucial to our understanding of these glassy systems.

Nevertheless, even for small systems, revealing completely their energy landscapes is difficult since they are high-dimensional functions. Existing approaches often omit some features of the landscapes for a feasible characterization. For instance, disconnectivity graphs (DG) connect attraction basins in the state space and show hierarchically how they are repeatedly segmented into smaller basins as energy decreases~\cite{becker1997topology}. DGs have been applied to analyze energy landscapes of systems from protein folding to machine learning~\cite{becker1997topology, zhou2009energy, ballard2017energy}, and can be improved using principal component analyses~\cite{komatsuzaki2005many}. However, DGs only show the segmentation into basins, without showing their entropy nor how states are exactly connected, especially as basins may have multiple instead of one connection to other states. Another common approach is multi-dimensional scaling (MDS), which aims to preserve the high-dimensional distance between two states in a plot of reduced dimension~\cite{mead1992review}. For instance, one may preserve the distance between pairs of states in one-dimensional plots~\cite{heuer1997properties}. Nevertheless, MDS only shows pairwise distance while dynamics on these systems are definitely more complex than pairwise interactions. 
%All these approaches attempt to characterize energy landscapes by omitting some of their features.

On the other hand, efforts have been made to reveal the relations among the ground states but omitting other parts of the energy landscape, which is a goal different from the present study. For instance, the hierarchical structure among ground states was revealed by examining the similarity in spin domains~\cite{hed2001spin, ciliberti2004quantitative, marinari2001equilibrium}. For constraint satisfiability problems, \cite{mezard2005clustering} showed that solutions are grouped into seperate clusters, while similarity among solutions are examined through relaxing discrete variables to be continuous during optimization~\cite{ercsey2011optimization}. Since this line of research primarily investigates ground states but not the rest of the energy landscape such as higher-lying local minima, and since the space spanned by the ground states is much smaller than the whole configurational space, they can investigate larger systems compared to the present study which aims to reveal the whole configurational space.

In this study, we introduce an approach to reveal for the complete energy landscape of complex disordered systems such as spin glasses and Boolean satisfiability problems. The approach is feasible on small systems, while for \red{larger} systems, we introduce a variant approach to obtain a partial energy landscape. \red{Here, we do not aim to contribute to the extensive studies in the so-called thermodynamics limit, i.e. systems with an infinite size, of spin glasses and Boolean satisfiability problems; instead, our goal is to offer insights on the similarity and difference between the behaviors of small systems and those theoretical predictions in the thermodynamic limit, by visualizing and analyzing the complete energy landscape in small systems, which is much less explored in existing studies.}

The obtained energy landscapes allow us to compute \red{semi-}analytically the \red{approximate} non-equilibrium dynamics at an arbitrary temperature for an arbitrary long time, out of reach by simulations limited by modern computational capability. In contrary to our common belief, we show that it can be less likely for the systems to attain ground states when temperature decreases, due to trapping in local minima; as time increases, trapping in individual minima ceases at different time, leading to multiple abrupt jumps in the ground-state probability. Our findings also provide insights on the effectiveness of simulated annealing compared with fixed-temperature dynamics, whereas only an extremely long annealing process that allows an escape from local minima may guarantee a ground state~\cite{bertsimas1993simulated}. All in all, our approach opens up a new platform for analyzing the non-equilibrium dynamics of glassy systems, and provides us with a clear, complete and new physical picture on their long-time behaviors inaccessible by existing approaches and numerics.

\section*{Methods}

We consider a system with $N$ Boolean variables $s_i=\pm 1$, such that $i=1,\dots,N$ and $\vs$ denotes the $N$-tuple $(s_1, s_2, \dots, s_N)$ representing a variable configuration. We then denote the energy or objective function of the system as $E(\vs)$. Here, we examine two glassy systems as examples, namely (i) spin glasses~\cite{mezard1987spin} and (ii) $K$-satisfiability problems~\cite{malik2009boolean}. \red{Here, we studied random instances; the difference between the energy lanscapes of random instances and those instances with planted solutions~\cite{barthel2002hiding, krzakala2012reweighted} is worth studying and will be studied elsewhere.}

For \emph{spin glasses}, each $\vs$ is a configuration of Ising spins and $E(\vs)$ is given by
\begin{equation}
	\label{eq_sgH}
	E(\vs)=\frac{1}{2}\sum_{i<j}a_{ij}J_{ij}s_i s_j,
\end{equation}
where $J_{ij}=+1$ with a probability $f_+$ and otherwise $J_{ij}=-1$; the adjacency matrix $a_{ij}=0,1$ characterizes different graph topologies. We multiply $E$ by a factor of $1/2$, such that a single spin flip leads to a unit change in energy. Depending on the topology, the parameter $f_+$ and the temperature, the spin system exhibits various phases such as paramagnetic, ferromagnetic and spin glass phases~\cite{mezard1987spin,sherrington1975solvable}.

For \emph{$K$-satisfiability problems}, or $K$-Sat for short, we introduce $M$ clauses of the form $(s_{\mu_1} \vee \overline{s}_{\mu_2}\vee \dots\vee s_{\mu_K})$ labeled by $\mu=1,\dots,M$, each with $K$ variables or their negation; the symbol $\vee$ corresponds to the ``or'' logical relation and the variables with an overline are negated. In this case, $E(\vs)$ is given by
\begin{equation}
	E(\vs)=\frac{1}{2^K}\sum_{\mu=1}^M\prod_{k=1}^K(1-J_{\mu, k}s_{\mu_k}),
\end{equation}
where randomly drawn $J_{\mu, k}=\pm 1$ corresponds to the presence of the original or the negated $k$-th variable in clause $\mu$. With the factor of $\frac{1}{2^K}$, each violated clause increases the energy by $1$ and the total energy is equivalent to the number of violated clauses. The ground state of the system is attained when $E=0$, i.e. all clauses are satisfied. Depending on the ratio $\alpha = M/N$, the system exhibits various phases including a satisfiable phase at small $\alpha$ with an algorithmic-easy and -hard regime, followed by an unsatisfiable phase at large $\alpha$~\cite{mezard2002random}. The phase transition between the ``easy'' and the ``hard'' phases is only well defined for systems in the thermodynamics limit, but not for small systems studied here; \red{\cut{nevertheless, algorithmic-easy and -hard regimes are still present in small systems as an $k$-Sat problem is more difficult to solve as $\alpha$ increases.} nevertheless, we show cases with $\alpha=1$ and $\alpha=4$ which correspond to the ``easy'' and the ``hard'' phases in the thermodynamics limit.}

Since there are $N$ Boolean variables in the above systems, there are $2^N$ different variable configurations. If we consider two configurations $\vs_a$ and $\vs_b$ to be connected in the configurational space if their hamming distance is $|\vs_a-\vs_b|=1$, i.e. they differ only in the state of a single variable, the configurational space is effectively an $N$-dimensional hypercube. 

\begin{figure}
	\centering
	\includegraphics[width=0.7\linewidth]{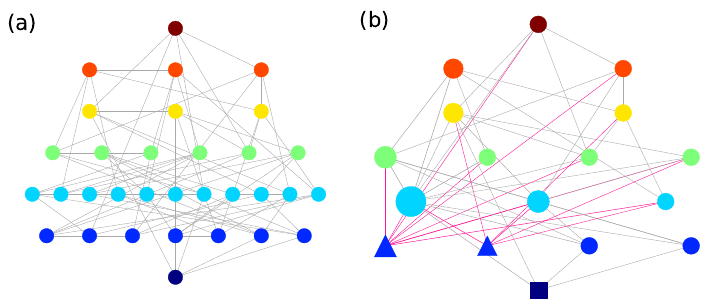}
	\vspace{-0,2cm}
	\caption{
		(a) An example of FEL with $2^N=32$ configurations from $E=7$ at the top to $E=0$ at the bottom, of a $3$-Sat problem with $N=5$ variables and $\alpha=4$. (b) The corresponding CEL with $\cC=17$ clusters. Global minima and local minima are shown in squares and triangles respectively; node size corresponds to the number of constituent configurations in the clusters; red links correspond to the connections to local minima.	}
	\label{fig_fel}
\end{figure}

To present this hypercube as an energy landscape, we take advantage of the integer disorders $J$ defined in the above systems, which lead to discrete energy levels. We then represent each variable configuration $\vs$ as a node in a network; two nodes are connected by a link if their hamming distance is 1. Next, we arrange nodes with the same energy on the same horizontal level in the network, with lower-energy configurations located at a lower row. We call this the \emph{full energy landscape} (FEL). For the sake of clear illustration, we show an example of FEL of a small $3$-Sat toy problem with $N=5$ and $\alpha=4$ in \fig{fig_fel}(a). One can see clearly how the $2^N=32$ different configurations are connected and arranged in different energy levels. Nevertheless, as $N$ further increases, the number of states increases exponentially and FELs quickly become computationally infeasible and difficult to be clearly visualized.

To simplify the energy landscape, we group connected nodes on the same energy level into \emph{clusters}; we denote $\cC$ to be the total number of clusters. Two clusters $a$ and $b$ are connected if any pair of their constituent variable configurations are connected; the weight $w_{ab}$ of the connection is the total number of links between their constituent configurations. We call this energy landscape the \emph{coarse-grained energy landscape} (CEL). The corresponding CEL of the FEL in \fig{fig_fel}(a) is shown in \fig{fig_fel}(b), where the number of nodes is reduced from $2^N=32$ in FEL to $\cC=17$ in CEL.

\red{One may remark that two configurations differ by a single spin or variable flip are usually considered indistinguishable in the theoretical analyses under the thermodynamics limit, in such case only energy barriers of a height of at least the order of $O(\sqrt{N})$ is relevant. Nevertheless, the objective of this study is not to analyze small systems as in the theoretical analyses in the thermodynamic limit, but instead we do the opposite to examine the trace of insights from the predictions in the thermodynamic limit to finite systems via visualizing the complete energy landscapes. In small systems, such a single spin or variable flip is important, for instance, in simulated annealing a single spin flip leading to higher or lower energy does affect its transition probability. Thus, in our case of small systems, we define configuration clusters using single variable flips.}

\section*{Results}
\subsection*{CELs with local minima}

We first analyze the CEL of small $K$-satisfiability problems with $K=3$ and $\alpha=4$, i.e. when the system is in the so-called ``Hard-Sat'' regime \red{in the thermodyanmics limit,} usually characterized by complex energy landscapes as we understand. Here we can only study systems up to a small size $N=25$, which may seem small compared to the studies on ground states~\cite{hed2001spin, ciliberti2004quantitative, marinari2001equilibrium}, but indeed the number of configurations visualizaed in our study is $2^{25} \approx 3.3\times 10^7$, which is much larger than the number of ground states, i.e. $O(10^2)$ or $O(10^3)$, analyzed in these studies.

\begin{figure}
	\centering
	\includegraphics[width=0.7\linewidth]{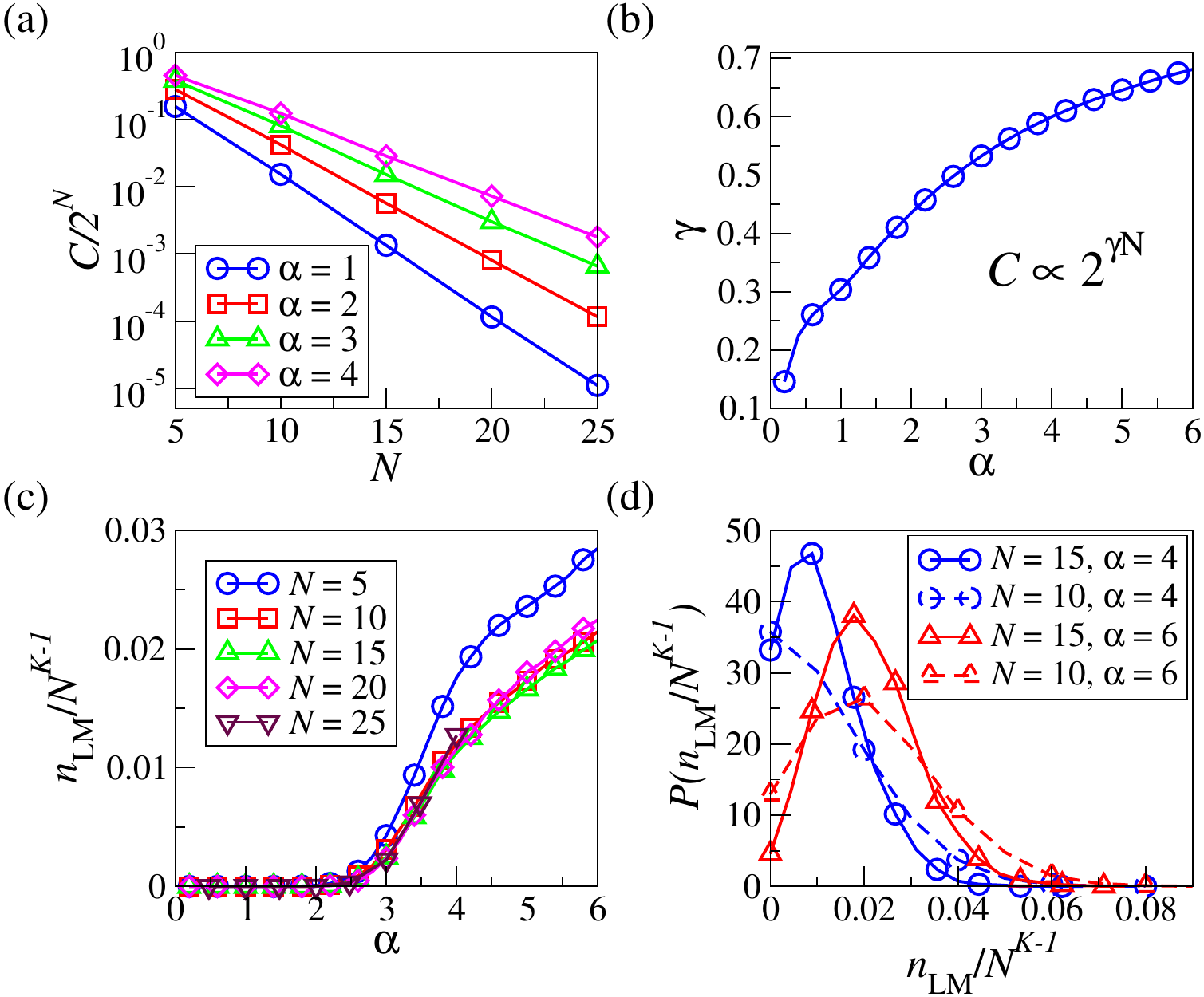}
	\vspace{-0,2cm}
	\caption{
		(a) The number of clusters in CEL divided by the number of configurations in FEL, i.e. $\cC/2^N$, as a function of $N$. (b) The exponent $\gamma$ in $\cC\propto2^{\gamma N}$ as a function of $\alpha$ in the $3$-Sat problem. (c) The number of local minima in CEL denoted as $n_{\rm LM}$, rescaled with $N^2$ of the $3$-Sat problem as a function of $\alpha$ for different system size $N$. (d) The rescaled probability distribution $P(n_{\rm LM}/N^{K-1})$. \red{All the results are obtained by averaging over 10000, 2000, 1000, 500, 100 realizations for cases with $N=5, 10, 15, 20$ and $25$ respectively.}
	}
	\label{fig_ncluster}
\end{figure}

As shown in \fig{fig_ncluster}(a), as $N$ further increases, the ratio of the number of clusters in CEL to the total number of variable configurations, i.e. $\cC/2^N$, decreases exponentially with $N$, implying that an extensive number of configurations can be grouped in CEL for a clear presentation. This also implies that $\cC\propto 2^{\gamma N}$, with $\gamma<1$. As shown in \fig{fig_ncluster}(b), $\gamma$ increases with $\alpha$, implying that the structure of energy landscape is more complicated at large $\alpha$, consistent with our understanding of algorithmic-hard regimes \red{in the theromdynamic limit,} compared with easy ones at small $\alpha$. One can also see that $\gamma$ approaches $\ln 2$ as $\alpha$ increases, implying that clusters are increasingly composed of individual nodes as there are more distinct energy levels, consistent with the segmentation of solution space in studies which focuses on analyzing ground states~\cite{krzakala2007gibbs}.

Interestingly, as shown in Fig. S1(a) of the \emph{Supplementary Information} (SI), the exponent $\gamma$ for $K$-Sat problems with different $K$ and $\alpha$ collapses onto a common function of $\alpha/K^2$. This implies that the decrease of nodes by grouping configurations in CEL is universal for different values of $N$, $M$ and $K$, which is further shown by the ratio $(\cC/2^N)^{1/(1-\gamma)}$ collapsed onto a common exponential decay against $N$ in Fig. S1(b). 
%This also implies that the decrease is more extensive with large $K$, since the satisfiability constraints are less restrictive in clauses with more variables and more configurations can be grouped in clusters because they are more likely to have the same energy. 

Other than a large reduction in the number of nodes, another advantage of CEL is the identification of local minima. Since connected configurations with the same energy are grouped in clusters in CELs, one can easily identify local minima as clusters where all neighbors are of higher energy; such identification is not trivial in FEL since it is difficult to examine if there exists a path from a configuration to a lower-energy one without passing through higher-energy configurations. In the CEL in \fig{fig_fel}(b), one can see that there exist two local minima (triangles) with $E=1$. 

We show in \fig{fig_cel}(a) the low-energy portion of another examplar CEL of spin glasses on random regular graphs (RRG) with $N=15$ and $f_+=0.5$; since the configurations $\vs$ and $-\vs$ have the same energy according to \req{eq_sgH}, one can observe a symmetric structure in the energy landscape as expected, including a pair of local minima at $E=3$. Another example of CEL of a $3$-Sat problem with $N=15$ and $\alpha=4$ is shown in \fig{fig_cel}(b), when the system is in the Hard-Sat regime, where six local minima are found at $E=1$. More examplar CELs of systems with larger $N$ are found in Fig.~S2 of the SI. In comparison, as shown in \fig{fig_cel}(c), the CEL of a $3$-Sat problem with $N=15$ and $\alpha=1$ \red{\cut{, i.e. when the system is in the ``easy-sat'' regime,}} is much simpler in structure without local minimum. This shows that the CELs \red{\cut{can reveal omprehensively} of small systems show similarity with} the expected structure of \red{the corresponding} energy landscapes in the ``hard'' and ``easy'' regimes \red{in the thermodynamics limit}.

In addition, CELs allow us to obtain the statistics of local minima, and the number of local minima $n_{\rm LM}$ is shown as a function of $\alpha$ for the $3$-Sat problem in \fig{fig_ncluster}(c). As we can see, local minima start to emerge beyond $\alpha\gtrsim 2.5$ and increase with $\alpha$. This is again consistent with the phenomenon of increasing algorithmic hardness as $\alpha$ increases. Interestingly, $n_{\rm LM}$ scales with $N^{K-1}$, which may imply that the emergence of local minima is related to the number of possible constraints per variable. We further show the distribution of $n_{\rm LM}/N^{K-1}$ in \fig{fig_ncluster}(d), where the distributions become narrower as $N$ increases. We remark that these results are different from most of the previous exhaustic studies on small combinatorial systems which mainly focus on ground states~\cite{ardelius2008exhaustive}.

%Although the largest system size of which CEL has been analyzed in \fig{fig_ncluster}(c) is $N=25$, we observe that the number of local minima rescaled with $1/N^2$ at different $N$ roughly collapse as a function of $\alpha$. 

We remark that CEL can be readily applied to systems with discrete coupling strength, which include many representative spin models and combinatorial optimization problems. The present approach can also accommodate discrete external fields which commensurate with the magnitude of conpuling strength. Nevertheless, for systems with non-discrete coupling strength and external fields, binning of energy would be a simple way to generalize our method to these systems.

\subsection*{Trapping Dynamics}

Thanks to the largely reduced number of nodes and the identification of local minima in CELs, they allow us to reveal the complete non-equilibrium dynamics when these glassy systems are trapped in local minima, at an arbitrary temperature for an arbitrarily long time. Here, one can formulate a matrix of transition probabilities $T_{a\to b}$ from a cluster $a$ to $b$, describing the Metropolis-Hasting (Markov Chain Monte Carlo (MCMC)) dynamics of the system following the Boltzmann distribution~\cite{metropolis1953equation,hastings1970monte}. In this case, $T_{a\to b}$ for $a\neq b$ is given by
\begin{equation}
	T_{a\to b}(\beta) = \frac{w_{ab}}{n_a N}\frac{e^{-\beta\Delta E_{a\to b}}}{e^{-\beta\Delta E_{a\to b}}+1}
\end{equation}
where $\Delta E_{a\to b} = E_{b}-E_{a}$ and $\beta$ is the inverse-temperature; $n_a$ corresponds to the size of cluster $a$, and $n_a N$ corresponds to the total number of links connecting its constituent configurations, including those internal links within cluster $a$. On the other hand, for the system to stay in cluster $a$, the system can either reject the transition to a configuration outside $a$ or transit to another configuration within $a$, with a total probability given by $T_{a\to a}(\beta)=1-\sum_{b\neq a} T_{a\to b}(\beta)$. We then denote the probabilities to find the system in configurations in individual clusters at time $t$ by a vector $\vP_t=(P_1,\dots,P_\cC)$, and express
\begin{equation}
	\label{eq_recursion_0}
	\vP_{t} = \cT_\beta\vP_{t-1} = \cT_\beta^t \vP_0,
\end{equation}
where $\cT_\beta$ is the transition matrix with element $T_{a\to b}(\beta)$. 

With the matrix $\cT_\beta$ for specific instances, one can conduct a spectral analysis to compute its eigenvalues, and we denote $\lambda_n(\beta)$ to be the $n$-th largest eigenvalue of $\cT_\beta$. \red{We will omit the dependence of $\lambda_n$ on $\beta$ in subsequent discussions for clarity. We first diagonalize $\cT_\beta$ as $\cT_\beta = \cQ_\beta \Lambda_\beta \cQ_\beta^{-1}$ by the diagonal matrix $\Lambda_\beta$ composed of the eigenvalues of $\cT_\beta$, and the matrix $\cQ$ composed of their corresponding eigenvectors. The power of $\cT_\beta$ can be computed by $\cT_\beta^t = \cQ_\beta \Lambda_\beta^t \cQ_\beta^{-1}$, such that $\vP_{t}$ in \req{eq_recursion_0} is given by}
\begin{equation}
	\label{eq_recursion}
	\red{  \vP_{t} =  \cQ_\beta \Lambda_\beta^t \cQ_\beta^{-1} \vP_0. }
\end{equation}
\red{We remark that $\Lambda_\beta^t$ can be readily computed by the power of the diagonal element of $\Lambda_\beta$ since it is a diagonal matrix. Thus, one can compute the probability of reaching any clusters in the CELs at any time step $t$. Moreover, by considering the limit $\Lambda^\infty_\beta = \lim_{t\rightarrow \infty} \Lambda_\beta^t$, one can evaluate the equilibrium probability $\vP_{t=\infty}$ of the system taking any configuration clusters after an infinitely long time.}

\red{We remark that finding the eigenvalues of $\cT_\beta$ not only facilitates the computation of $\vP_{t}$, but also provides important insights into the structure of the energy landscape and its non-equilibrium dynamics. Since $\cT$ is a transitional probability matrix, by Perron–Frobenius theorem~\cite{meyer2023matrix}, the absolute values of its eigenvalues are bounded by 1, and except those eigenvalues equal to 1, all of them vinish as $t$ is large. Therefore, the closeness of an eigenvalue to 1 thus gives us a measure of the metastability the corresponding eigenmode, e.g. trapping in a local minima, and thus how strong the non-equilibrium dynamics is trapped.}

\begin{figure}
	\centering
	\includegraphics[width=0.5\linewidth]{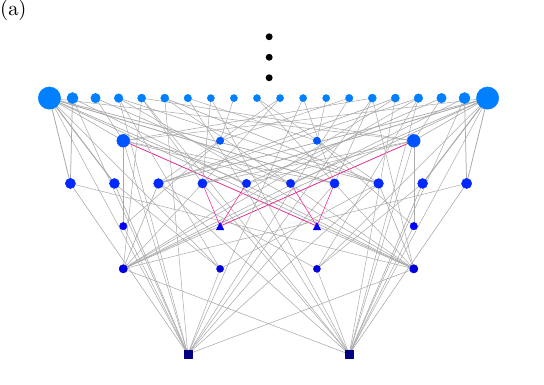}
	\includegraphics[width=0.5\linewidth]{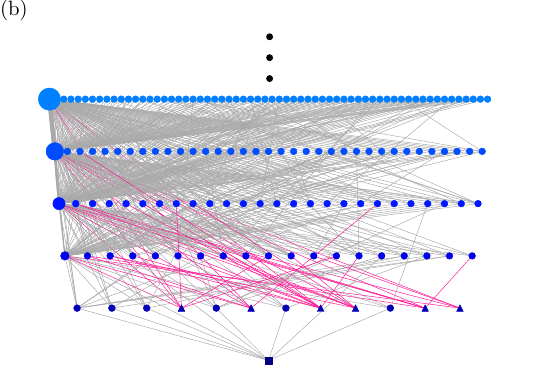}
	\includegraphics[width=0.5\linewidth]{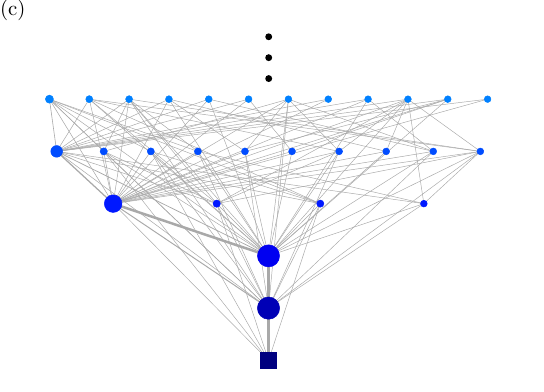}
	\vspace{-0,5cm}
	\caption{
		The low-energy portion of examplar CELs for an instance of (a) spin glass on random regular graph with $f_+=0.5$, and 3-Sat problems with (b) $\alpha=4$ and (c) $\alpha=1$, all with $N=15$. Global minima and local minima are shown in squares and triangles respectively; node size corresponds to the number of constituent configurations in the clusters; red links correspond to the connections to local minima.
	}
	\label{fig_cel}
\end{figure}

\begin{figure*}
	\centering
	\includegraphics[width=\linewidth]{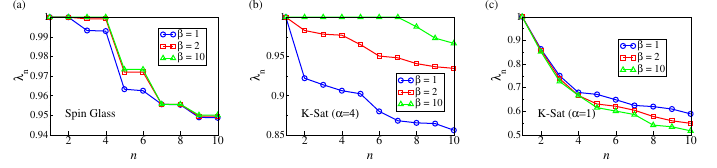}
	\vspace{-0,2cm}
	\caption{
		The first 10 largest eigenvalues, i.e. $\lambda_n(\beta)$ with $n=1,\dots,10$, of the transition matrix $\cT_\beta$ at $\beta=1,2,10$, for the examplar spin glass, and the \red{\cut{``hard'' and ``easy''}} 3-Sat problem instances \red{with $\alpha=4$ and $\alpha=1$,} of which CELs are shown in \fig{fig_cel}(a) to (c) respectively.}
	\label{fig_eigenvalue}
\end{figure*}

%With the matrix $\cT_\beta$ for specific instances, one can conduct a spectral analysis to compute its eigenvalues. We first denote $\lambda_n(\beta)$ to be the $n$-th largest eigenvalue of $\cT_\beta$.
%; 
\red{As an example}, we show in \fig{fig_eigenvalue}(a) to (c) $\lambda_1$ to $\lambda_{10}$ at different inverse-temperatures $\beta$ for the spin glass, \red{\cut{``hard'' and ``easy''}} $K$-Sat instances \red{with $\alpha=4$ and $\alpha=1$} shown in \fig{fig_cel}(a) to (c) respectively. In \fig{fig_eigenvalue}(a), we first note that the eigenvalues of the spin glass instance are in pairs due to the symmetric nature of its energy landscape. More interestingly, as shown in both \fig{fig_eigenvalue}(a) and (b), $\lambda_n$ differ more at small $\beta$, but the few eigenvalues after $\lambda_1$ start to approach 1 as $\beta$ increases. The number of eigenvalues approaching 1 is equal to the number of local minima in the corresponding CELs, i.e. $\lambda_3$ and $\lambda_4$ of the spin glass instances correspond to the two symmetric local minima in \fig{fig_cel}(a) and $\lambda_2$ to $\lambda_7$ of the \red{\cut{``hard''}} $3$-Sat instance \red{with $\alpha=4$} correspond to the six local minima with $E=1$ in \fig{fig_cel}(b). The increasing proximity of these eigenvalues to 1 also corresponds to an increasing trapping in local minima when $\beta$ increases, comparable to the trapping in global minima with $\lambda_1=1$. This also raises a question on whether the systems equilibrate at the ground states at zero temperature (i.e. $\beta\to\infty$), since $\lambda_n\to 1$ for the local minima and are equivalent to $\lambda_1=1$ at the ground states. In comparison, as shown in \fig{fig_eigenvalue}(c), other than the largest eigenvalues $\lambda_1$, the other eigenvalues of the \red{\cut{``easy''}} $3$-Sat instance \red{($\alpha=1$)} with CEL shown in \fig{fig_cel}(c) do not approach $1$ when $\beta$ increases. \red{Instead, these eigenvalues decrease as $\beta$ increases. This suggests that these $\lambda^t_n$'s vanish faster as $t$ grows, implying that the system converges to the global faster at a lower temperature.}

\begin{figure*}
	\centering
	\includegraphics[width=\linewidth]{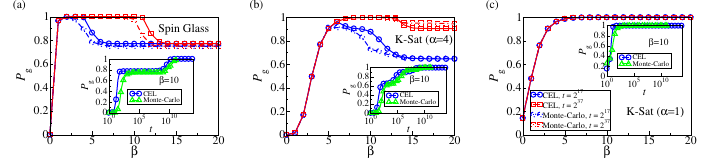}
	\vspace{-0,2cm}
	\caption{
		The probability $\pg$ of finding the ground states in the examplar spin glass, and the \red{\cut{``hard'' and ``easy''}} 3-Sat problem instances \red{with $\alpha=4$ and $\alpha=1$,} of which CELs are shown in \fig{fig_cel}(a) to (c) respectively, as a function of inverse-temperature $\beta$ obtained by \req{eq_recursion} after \red{$t=2^{17}$ and $2^{37}$} iterations, compared with simulation results. Insets: $\pg$ as a function of time $t$.	}
	\label{fig_dynamics}
\end{figure*}

Since we can obtain the complete transition matrix for these small systems through CELs, one can compute their complete dynamics at an arbitrary temperature for an arbitrarily long time using \req{eq_recursion}. Starting with a uniform $\vP_0$, we show the probability $\pg$ of the spin glass, and \red{\cut{the ``hard'' and ``easy''}} $3$-Sat instances \red{with $\alpha=4$ and $\alpha=1$} being in the ground state after \red{$t=2^{17}$} (blue circles) and \red{$2^{37}$} (red squares) iteration steps as a function of $\beta$ in \fig{fig_dynamics}. As we can see in \fig{fig_dynamics}(a) and (b), for both spin glass and \red{\cut{``hard''}} $3$-Sat instances \red{with $\alpha=4$}, $\pg$ first increases with $\beta$ as expected, but remarkably decreases as $\beta$ further increases; the MCMC simulation results are in good agreement with these theoretical predictions by CELs. These results imply that with a random initial condition, the trapping at local minima becomes more significant as temperature decreases below some specific values and it is less likely to locate the ground states within a finite time. Such observations have a strong implication on the cooling procedure in physics-inspired optimization algorithms such as simulated annealing. In comparison, instead of an optimal non-zero temperature, $\pg$ for the \red{\cut{``easy''}} $3$-Sat instance \red{with $\alpha=1$} monotonically increases with $\beta$, i.e. monotonically decreases with temperature, implying a zero temperature maximizes the probability of locating the ground state as expected.

As we can see, time $t$ seems to play an important role as the optimal range of $\beta$ with high $\pg$ widens with $t$. We further show in the insets of \fig{fig_dynamics} that $\pg$ does increase with $t$. Nevertheless, for spin glass and the ``hard'' $3$-Sat instance as shown in the insets of \fig{fig_dynamics}(a) and (b) respectively, the increase of $\pg$ is not smooth and multiple jumps and plateaus are observed, implying that local minima are completing with the global minima for the probability but they cease to trap the system at different time $t$. This phenomenon can be explained by eigenvalues, where a sufficiently large $t$ makes $\lambda_n^t$ of the local minima sufficiently less than 1, despite $\lambda_n\approx 1$ (see again \fig{fig_eigenvalue}(a) and (b)). This also implies that for glassy systems such as spin glasses and ``hard'' $K$-Sat problems, the proximity of $\lambda_n$ to 1 is related to the ability or stiffness of individual minima in trapping the system, which may depend on their entropy or number of external connections; one may thus estimate the characteristic duration of trapping in individual minima using $\lambda_n$. In comparison, for the \red{\cut{``easy''}} $3$-Sat instance \red{with $\alpha=1$} as shown in the inset of \fig{fig_dynamics}(c), $\pg$ increases smoothly with time $t$, implying there is no trapping in dynamics.

As the eigenvalue $\lambda_n$ of local minima approaches but is not equal to 1, one may anticipate that utlimately all $\lambda_n^t$ of the local minima would be sufficiently less than 1, and only the gobal minima left. In otherwords, the MCMC dynamics ultimately lead to the ground states. However, in a practically shorter time-scale, we see that \emph{spontaneous ergodicity breaking}~\cite{marinari1996numerical, cugliandolo1995weak} emerges as the time-scale is sufficiently large for the systems to explore the whole configurational space according to the equilibrium distribution.

We remark that MCMC simulations are in good agreement with theoretical predictions, including the drop in $\pg$ with $\beta$ and the abrupt jumps and plateaus of $\pg$ at small $t$ in \fig{fig_dynamics}(a) and (b), while the small discrepancies may come from the mean-field nature of the clustered transition probabilities in \req{eq_recursion}. In addition, since one can easily compute $\cT_\beta^t$ for an arbitrarily large $t$, e.g. $10^{14}$ in the insets of \fig{fig_dynamics}, by repeatedly powering $\cT_\beta^t$ and its products, one can obtain the long-time dynamics by \req{eq_recursion} out of reach by modern computational capability. For the sake of a clear illustration and elaboration, we show the above results for only three instances; in Fig.~S3 of the SI, we show that the sample-averaged $\pg$ \red{with 200 samples }exhibits a similar behavior against $\beta$ and $t$. Sample-averaged MCMC simulation results are also in good agreement with theoretical predictions.

\subsection*{Implications on \red{\cut{Simulated Annealing}Cooling}}

The eigenvalues of the transition matrix $\cT_\beta$ from CELs also provide implications on the essence of cooling in \red{identifying low-lying states, e.g.} simulated annealing (SA), especially for glassy systems such as spin glasses and ``hard $K$-Sat instances. As we see from \fig{fig_eigenvalue}(a) and (b), the difference among eigenvalues is large at small $\beta$ when the system can distinguish $\lambda_1$ which attributes to the global minima from $\lambda_n$ which attributes to the local minima. \red{\cut{With random initial conditions, }}As $\beta$ increases, these largest eigenvalues are getting closer in values. By cooling the system from a high temperature\red{\cut{ in SA}}, the system does not start with a random state at the beginning of each cooling stage but instead continuously biases towards the global minima due to difference between its $\lambda_1$ from other $\lambda_n$, though this difference is vanishing. This suggests that \red{\cut{SA}cooling} is more effective in identifying ground states compared with fixed-temperature dynamics. Nevertheless, once the system is trapped in a local minimum, lowering temperature in SA does not help the system escape from the minimum, and only an extremely slow (and potentially infeasible) cooling schedule may help.

\subsection*{Partial Coarse-grained Energy Landscape (PCEL)}

The computation of CELs is only feasible for systems with small size $N$ since it requires examining all $2^N$ variable configurations. Nevertheless, for large systems, we introduce a method to obtain a \emph{partial coarse-grained energy landscape} (PCEL) for the low-energy configurations. In this case, we sample variable configurations by MCMC simulations at a fixed \emph{sampling inverse-temperature} $\bs$ for $T$ steps, and restart the sampling with random initial conditions for multiple times. We record all the sampled configurations for the construction of PCELs following the same procedures as in CELs. By using an appropriate $\bs$, one can extract specific part of the energy landscape, for instance, a moderately large $\bs$ for extracting the low-energy configurations.

We remark that PCELs are only approximations since MCMC simulations with finite time $T$ can only sample a small fraction of all $2^N$ configurations, though the number of sampled configurations for systems with large $N$ can be significantly larger than those of the small systems we presented before. In addition, it can happen that clusters with the same energy in PCELs indeed belong to a larger cluster since not all the intermediate configurations between the two clusters are sampled. An example of PCEL for a $K$-Sat problem with $N=50$ is shown in Fig.~S4(a) of the SI. Since we are mainly interested in the glassy behaviors contributed by the global and local minima, to further simplify the analyses, we make one more approximation to leave only a single shortest path between minima in PCELs to obtain a simplified transition matrix $\tilde\cT_\beta$; the simplified version of PCEL in Fig.~S4(a) is shown in Fig.~S4(b). We found that the results obtained by the simplified $\tilde\cT_\beta$ are similar to those without this simplification.

\begin{figure}
	\centering
	\includegraphics[width=0.6\linewidth]{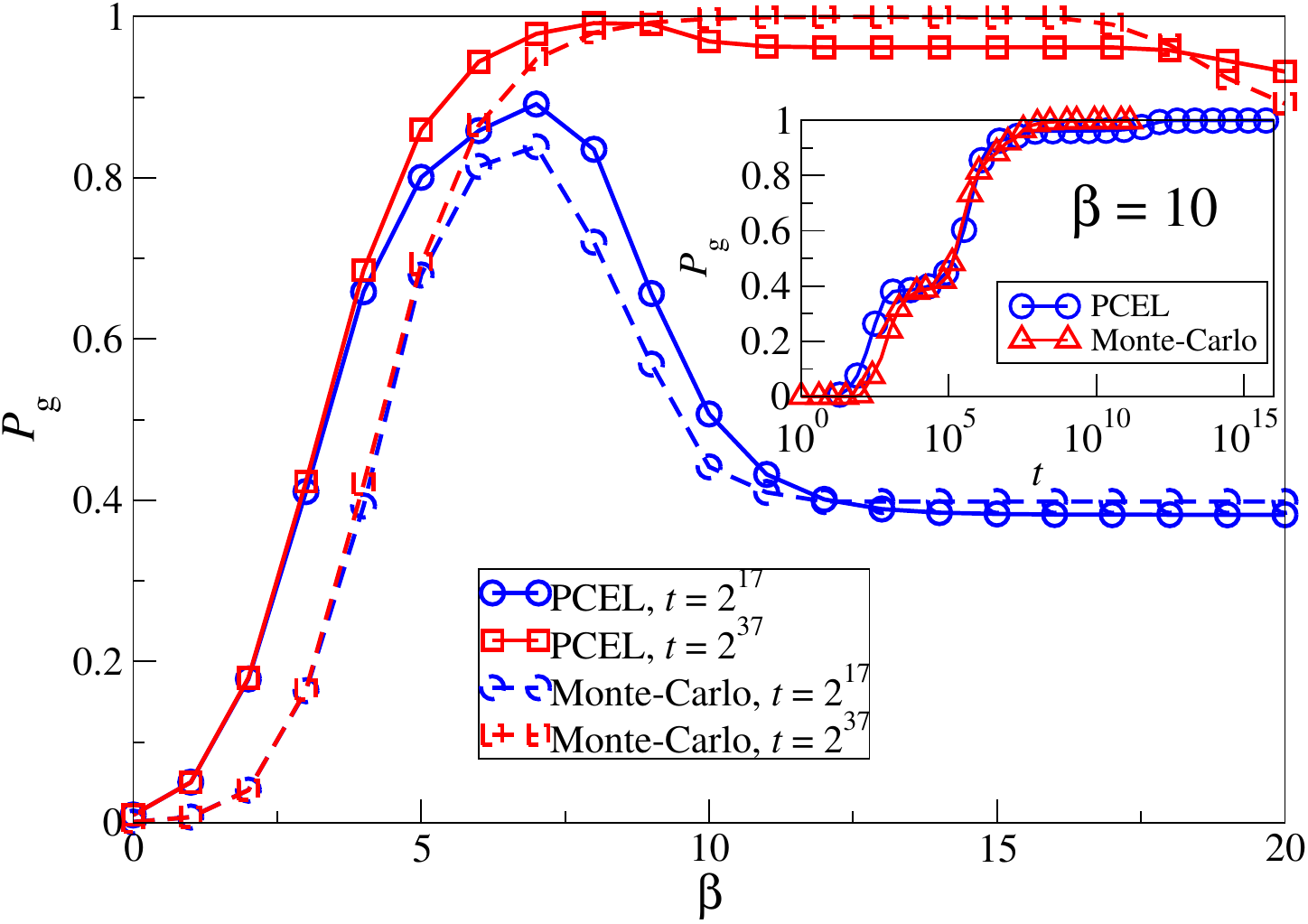}
	\vspace{-0,2cm}
	\caption{
		The probability $\pg$ for a $3$-Sat problem instance with $N=50$ of which PCELs are shown in Fig.~S4, as a function of $\beta$, obtained by the transition matrix from PCELs sampled for $T=10^5$ steps at $\bs=5$ with $10$ re-starts, and then by \req{eq_recursion} after \red{$t=2^{17}$ and $2^{37}$} iterations, compared with simulation results. Insets: $\pg$ as a function of time $t$, where a factor of $\ln[2^N/n(\vs_{\rm sampled})]$ has been multiplied to $t$ in the results obtained by PCELs since only part of the energy landscape is extracted, with $n(\vs_{\rm sampled})$ to be the number of sampled configurations.
	}
	\label{fig_largeN}
\end{figure}

The major advantage in using PCELs to analyze system dynamics is that a single MCMC procedure to extract the PCEL at a single $\bs$ can provide us with the dynamics of the system at an arbitrary temperature for an arbitrarily long time out of which by simulations. We show the dynamics of a $3$-Sat problem with $N=50$ in \fig{fig_largeN},
which is obtained by the simplified PCEL shown in Fig.~S4(b) sampled at a single $\bs=5$. The theoretical results agree well with simulations at different $\beta$ except those at small $\beta$ when the system explores high-energy configurations while PCELs focus on low-energy configurations. The corresponding sampled-averaged plot is shown in Fig.~S5. The same phenomena as in the small systems are observed, namely the drop in $\pg$ as temperature decreases, as well as the jumps in $\pg$ as time $t$ increases. These results suggest that the findings based on CELs in small systems are also observed in large systems, which show that CELs and PCELs open up a new platform for us to reveal the long-time non-equilibrium dynamics for glassy systems.

\section*{Conclusion}
We introduced an approach called Coarse-grained Energy Landscape (CEL) to reveal for the complete energy landscapes of small glassy and non-glassy systems, showing clearly their differences. In terms of methodology, by formulating CELs and analyzing their transition matrix, one can analytically compute the non-equilibrium dynamics of a system at an arbitrary temperature for an arbitrary long time, out of reach by existing theories and numerics, and again revealing clearly the differences between glassy and non-glassy systems. For large systems, we introduce a variant approach to partially reveal the energy landscapes, which allow us to conduct the same analysis as in small systems. In terms of understanding, we show a clear and complete physical picture on how glassy systems are trapped in local minima, as evident from the drop in the ground-state probability as temperature decreases as well as their abrupt jumps as time increases. Such phenomena are not observed in non-glassy systems. Simulation results agree well with theoretical predictions. Our work advances methodology by a new tool for analyzing the non-equilibrium dynamics of complex disordered systems, which generate clear, complete and new understandings and insights on their long-time behavior inaccessible by existing approaches.

%%%%%%%%%%%% fig 1 %%%%%%%%%%%%
%%%%%%%%%%%%%%%%%%%%%%%%%%

%%%%%%%%%%%% fig 2 %%%%%%%%%%%%
%%%%%%%%%%%%%%%%%%%%%%%%%%

%%%%%%%%%%%% fig 3 %%%%%%%%%%%%
%%%%%%%%%%%%%%%%%%%%%%%%%%

%%%%%%%%%%%% fig 4 %%%%%%%%%%%%
%%%%%%%%%%%%%%%%%%%%%%%%%%

%%%%%%%%%%%% fig 5 %%%%%%%%%%%%
%%%%%%%%%%%%%%%%%%%%%%%%%%

%%%%%%%%%%%% fig 6 %%%%%%%%%%%%
%%%%%%%%%%%%%%%%%%%%%%%%%%

\section*{Acknowledgments}
This work is fully supported by the Research Grants Council of the Hong Kong Special Administrative Region, China (Projects No. GRF 18304316, GRF 18301217, GRF 18301119 and GRF 18300623), the Dean's Research Fund of the Faculty of Liberal Arts and Social Sciences, The Education University of Hong Kong, Hong Kong Special Administrative Region, China (Projects No: FLASS/DRF 04418, FLASS/ROP 04396 and FLASS/DRF 04624), and the Research Development Office Internal Research Grant, The Education University of Hong Kong, Hong Kong Special Administrative Region, China (Projects No. RG67 2018-2019R R4015 and No. RG31 2020-2021R R4152).

\section*{Author contributions statement}
H.F.P and C.H.Y designed research, derived algorithm, created synthetic data, analyzed data and wrote the paper.

\section*{Data Availability}
The codes used for the analyses during the current study are available from the corresponding author on reasonable request.


\section*{Reference}
\begin{thebibliography}{10}
\urlstyle{rm}
\expandafter\ifx\csname url\endcsname\relax
  \def\url#1{\texttt{#1}}\fi
\expandafter\ifx\csname urlprefix\endcsname\relax\def\urlprefix{URL }\fi
\expandafter\ifx\csname doiprefix\endcsname\relax\def\doiprefix{DOI: }\fi
\providecommand{\bibinfo}[2]{#2}
\providecommand{\eprint}[2][]{\url{#2}}

\bibitem{mezard1987spin}
\bibinfo{author}{M{\'e}zard, M.}, \bibinfo{author}{Parisi, G.} \&
  \bibinfo{author}{Virasoro, M.~A.}
\newblock \emph{\bibinfo{title}{Spin glass theory and beyond: An Introduction
  to the Replica Method and Its Applications}}, vol.~\bibinfo{volume}{9}
  (\bibinfo{publisher}{World Scientific Publishing Company},
  \bibinfo{year}{1987}).

\bibitem{nishimori2001statistical}
\bibinfo{author}{Nishimori, H.}
\newblock \emph{\bibinfo{title}{Statistical physics of spin glasses and
  information processing: an introduction}}.
\newblock \bibinfo{number}{111} (\bibinfo{publisher}{Clarendon Press},
  \bibinfo{year}{2001}).

\bibitem{krzakala2007gibbs}
\bibinfo{author}{Krzaka{\l}a, F.}, \bibinfo{author}{Montanari, A.},
  \bibinfo{author}{Ricci-Tersenghi, F.}, \bibinfo{author}{Semerjian, G.} \&
  \bibinfo{author}{Zdeborov{\'a}, L.}
\newblock \bibinfo{journal}{\bibinfo{title}{Gibbs states and the set of
  solutions of random constraint satisfaction problems}}.
\newblock {\emph{\JournalTitle{Proceedings of the National Academy of
  Sciences}}} \textbf{\bibinfo{volume}{104}}, \bibinfo{pages}{10318--10323}
  (\bibinfo{year}{2007}).

\bibitem{becker1997topology}
\bibinfo{author}{Becker, O.~M.} \& \bibinfo{author}{Karplus, M.}
\newblock \bibinfo{journal}{\bibinfo{title}{The topology of multidimensional
  potential energy surfaces: Theory and application to peptide structure and
  kinetics}}.
\newblock {\emph{\JournalTitle{The Journal of chemical physics}}}
  \textbf{\bibinfo{volume}{106}}, \bibinfo{pages}{1495--1517}
  (\bibinfo{year}{1997}).

\bibitem{zhou2009energy}
\bibinfo{author}{Zhou, Q.} \& \bibinfo{author}{Wong, W.~H.}
\newblock \bibinfo{journal}{\bibinfo{title}{Energy landscape of a spin-glass
  model: exploration and characterization}}.
\newblock {\emph{\JournalTitle{Physical Review E}}}
  \textbf{\bibinfo{volume}{79}}, \bibinfo{pages}{051117}
  (\bibinfo{year}{2009}).

\bibitem{ballard2017energy}
\bibinfo{author}{Ballard, A.~J.} \emph{et~al.}
\newblock \bibinfo{journal}{\bibinfo{title}{Energy landscapes for machine
  learning}}.
\newblock {\emph{\JournalTitle{Physical Chemistry Chemical Physics}}}
  \textbf{\bibinfo{volume}{19}}, \bibinfo{pages}{12585--12603}
  (\bibinfo{year}{2017}).

\bibitem{komatsuzaki2005many}
\bibinfo{author}{Komatsuzaki, T.} \emph{et~al.}
\newblock \bibinfo{journal}{\bibinfo{title}{How many dimensions are required to
  approximate the potential energy landscape of a model protein?}}
\newblock {\emph{\JournalTitle{The Journal of chemical physics}}}
  \textbf{\bibinfo{volume}{122}}, \bibinfo{pages}{084714}
  (\bibinfo{year}{2005}).

\bibitem{mead1992review}
\bibinfo{author}{Mead, A.}
\newblock \bibinfo{journal}{\bibinfo{title}{Review of the development of
  multidimensional scaling methods}}.
\newblock {\emph{\JournalTitle{Journal of the Royal Statistical Society: Series
  D (The Statistician)}}} \textbf{\bibinfo{volume}{41}},
  \bibinfo{pages}{27--39} (\bibinfo{year}{1992}).

\bibitem{heuer1997properties}
\bibinfo{author}{Heuer, A.}
\newblock \bibinfo{journal}{\bibinfo{title}{Properties of a glass-forming
  system as derived from its potential energy landscape}}.
\newblock {\emph{\JournalTitle{Physical review letters}}}
  \textbf{\bibinfo{volume}{78}}, \bibinfo{pages}{4051} (\bibinfo{year}{1997}).

\bibitem{hed2001spin}
\bibinfo{author}{Hed, G.}, \bibinfo{author}{Hartmann, A.~K.},
  \bibinfo{author}{Stauffer, D.} \& \bibinfo{author}{Domany, E.}
\newblock \bibinfo{journal}{\bibinfo{title}{Spin domains generate hierarchical
  ground state structure in j=$\pm$1 spin glasses}}.
\newblock {\emph{\JournalTitle{Physical Review Letters}}}
  \textbf{\bibinfo{volume}{86}}, \bibinfo{pages}{3148} (\bibinfo{year}{2001}).

\bibitem{ciliberti2004quantitative}
\bibinfo{author}{Ciliberti, S.} \& \bibinfo{author}{Marinari, E.}
\newblock \bibinfo{journal}{\bibinfo{title}{A quantitative clustering approach
  to ultrametricity in spin glasses}}.
\newblock {\emph{\JournalTitle{Journal of statistical physics}}}
  \textbf{\bibinfo{volume}{115}}, \bibinfo{pages}{557--580}
  (\bibinfo{year}{2004}).

\bibitem{marinari2001equilibrium}
\bibinfo{author}{Marinari, E.}, \bibinfo{author}{Martin, O.~C.} \&
  \bibinfo{author}{Zuliani, F.}
\newblock \bibinfo{journal}{\bibinfo{title}{Equilibrium valleys in spin glasses
  at low temperature}}.
\newblock {\emph{\JournalTitle{Physical Review B}}}
  \textbf{\bibinfo{volume}{64}}, \bibinfo{pages}{184413}
  (\bibinfo{year}{2001}).

\bibitem{mezard2005clustering}
\bibinfo{author}{M{\'e}zard, M.}, \bibinfo{author}{Mora, T.} \&
  \bibinfo{author}{Zecchina, R.}
\newblock \bibinfo{journal}{\bibinfo{title}{Clustering of solutions in the
  random satisfiability problem}}.
\newblock {\emph{\JournalTitle{Physical Review Letters}}}
  \textbf{\bibinfo{volume}{94}}, \bibinfo{pages}{197205}
  (\bibinfo{year}{2005}).

\bibitem{ercsey2011optimization}
\bibinfo{author}{Ercsey-Ravasz, M.} \& \bibinfo{author}{Toroczkai, Z.}
\newblock \bibinfo{journal}{\bibinfo{title}{Optimization hardness as transient
  chaos in an analog approach to constraint satisfaction}}.
\newblock {\emph{\JournalTitle{Nature Physics}}} \textbf{\bibinfo{volume}{7}},
  \bibinfo{pages}{966--970} (\bibinfo{year}{2011}).

\bibitem{bertsimas1993simulated}
\bibinfo{author}{Bertsimas, D.}, \bibinfo{author}{Tsitsiklis, J.} \emph{et~al.}
\newblock \bibinfo{journal}{\bibinfo{title}{Simulated annealing}}.
\newblock {\emph{\JournalTitle{Statistical science}}}
  \textbf{\bibinfo{volume}{8}}, \bibinfo{pages}{10--15} (\bibinfo{year}{1993}).

\bibitem{malik2009boolean}
\bibinfo{author}{Malik, S.} \& \bibinfo{author}{Zhang, L.}
\newblock \bibinfo{journal}{\bibinfo{title}{Boolean satisfiability from
  theoretical hardness to practical success}}.
\newblock {\emph{\JournalTitle{Communications of the ACM}}}
  \textbf{\bibinfo{volume}{52}}, \bibinfo{pages}{76--82}
  (\bibinfo{year}{2009}).
  
  \bibitem{barthel2002hiding}krzakala2012reweighted
  Wolfgang Barthel, Alexander K Hartmann, Michele Leone, Federico Ricci-Tersenghi, Martin Weigt, and Riccardo Zecchina.
  \textit{Hiding solutions in random satisfiability problems: A statistical mechanics approach}.
  Physical Review Letters, 88(18):188701, 2002.
  
  \bibitem{krzakala2012reweighted}
  Florent Krzakala, Marc Mézard, and Lenka Zdeborová.
  \textit{Reweighted belief propagation and quiet planting for random k-sat}.
  Journal on Satisfiability, Boolean Modeling and Computation, 8(3-4):149--171, 2012.

\bibitem{sherrington1975solvable}
\bibinfo{author}{Sherrington, D.} \& \bibinfo{author}{Kirkpatrick, S.}
\newblock \bibinfo{journal}{\bibinfo{title}{Solvable model of a spin-glass}}.
\newblock {\emph{\JournalTitle{Physical review letters}}}
  \textbf{\bibinfo{volume}{35}}, \bibinfo{pages}{1792} (\bibinfo{year}{1975}).

\bibitem{mezard2002random}
\bibinfo{author}{M{\'e}zard, M.} \& \bibinfo{author}{Zecchina, R.}
\newblock \bibinfo{journal}{\bibinfo{title}{Random k-satisfiability problem:
  From an analytic solution to an efficient algorithm}}.
\newblock {\emph{\JournalTitle{Physical Review E}}}
  \textbf{\bibinfo{volume}{66}}, \bibinfo{pages}{056126}
  (\bibinfo{year}{2002}).

\bibitem{ardelius2008exhaustive}
\bibinfo{author}{Ardelius, J.} \& \bibinfo{author}{Zdeborov{\'a}, L.}
\newblock \bibinfo{journal}{\bibinfo{title}{Exhaustive enumeration unveils
  clustering and freezing in the random 3-satisfiability problem}}.
\newblock {\emph{\JournalTitle{Physical Review E}}}
  \textbf{\bibinfo{volume}{78}}, \bibinfo{pages}{040101}
  (\bibinfo{year}{2008}).

\bibitem{metropolis1953equation}
\bibinfo{author}{Metropolis, N.}, \bibinfo{author}{Rosenbluth, A.~W.},
  \bibinfo{author}{Rosenbluth, M.~N.}, \bibinfo{author}{Teller, A.~H.} \&
  \bibinfo{author}{Teller, E.}
\newblock \bibinfo{journal}{\bibinfo{title}{Equation of state calculations by
  fast computing machines}}.
\newblock {\emph{\JournalTitle{The journal of chemical physics}}}
  \textbf{\bibinfo{volume}{21}}, \bibinfo{pages}{1087--1092}
  (\bibinfo{year}{1953}).

\bibitem{hastings1970monte}
\bibinfo{author}{Hastings, W.~K.}
\newblock \emph{\bibinfo{title}{Monte Carlo sampling methods using Markov
  chains and their applications}} (\bibinfo{publisher}{Oxford University
  Press}, \bibinfo{year}{1970}).

\bibitem{meyer2023matrix}
Carl D Meyer.
\textit{Matrix analysis and applied linear algebra}.
SIAM, 2023.

\bibitem{marinari1996numerical}
\bibinfo{author}{Marinari, E.}, \bibinfo{author}{Parisi, G.},
  \bibinfo{author}{Ruiz-Lorenzo, J.} \& \bibinfo{author}{Ritort, F.}
\newblock \bibinfo{journal}{\bibinfo{title}{Numerical evidence for
  spontaneously broken replica symmetry in 3d spin glasses}}.
\newblock {\emph{\JournalTitle{Physical review letters}}}
  \textbf{\bibinfo{volume}{76}}, \bibinfo{pages}{843} (\bibinfo{year}{1996}).

\bibitem{cugliandolo1995weak}
\bibinfo{author}{Cugliandolo, L.~F.} \& \bibinfo{author}{Kurchan, J.}
\newblock \bibinfo{journal}{\bibinfo{title}{Weak ergodicity breaking in
  mean-field spin-glass models}}.
\newblock {\emph{\JournalTitle{Philosophical Magazine B}}}
  \textbf{\bibinfo{volume}{71}}, \bibinfo{pages}{501--514}
  (\bibinfo{year}{1995}).

\end{thebibliography}
\end{document}